\documentclass[a4paper]{jpconf}
\usepackage{graphicx}
\begin{document}
\title{Ionic channels as electrostatic amplifiers of  charge fluctuations}

\author{D.G. Luchinsky$^1$, R. Tindjong$^1$, I. Kaufman$^2$, P.V.E. McClintock$^1$ and R.S. Eisenberg$^3$}

\address{$^1$Department of physics, Lancaster University, Lancaster, LA1 4YB, UK}
\address{$^2$The Russian Research Institute for Metrological Service,
\\ Gosstandart, Moscow, 119361, Russia}
\address{$^3$Department of Molecular Biophysics and Physiology, Rush Medical College,
\\ 1750 West Harrison, Chicago, IL 60612, USA;}

\ead{r.tindjong@lancaster.ac.uk}

\begin{abstract}
Electrostatic interaction between ions in an ionic channel and the
charge fluctuations in the channel mouth are considered. It is shown
that the charge fluctuations can be enhanced in the channels with
low dielectric constant and result in strong modulation of the
potential barrier at the selectivity site. The effect of charge
fluctuational on transition probabilities in other molecular
dynamical systems is briefly discussed.
\end{abstract}

\section{Introduction}
\label{s:introduction}

Electrostatic interactions and fluctuations play crucial role in
controlling transition probabilities, conduction and selectivity in
nanoscale molecular systems. It is also well-know that the
long-range Coulomb interaction leads to the strong correlation of
ions motion and is responsible for the ions distribution and the
energy fluctuations in electrolytes~\cite{Debye:23,Landau:80a}.
However, in traditional approach to the calculations of the escape
rates in molecular biology the charge fluctuations in the bulk are
usually neglected~\cite{Hille:92}. In this paper the effect of
charge fluctuations on the probabilities of transition of ions
through an open ionic channel is considered.

Specifically, we consider ion transport across lipid membrane with
small dielectric constant. The motion is described as a
self-consistent solution of the coupled Poisson and Langevin
equations for ions moving in an open ionic
channel~\cite{Chung:98,Chung:99}. The distribution of ions arrival
time and the charge fluctuations at the channel mouth are
investigated numerically and compared to the theoretical
predictions. Next, the electrostatic energy profile of one ion in an
aqueous channel through a lipid membrane charged at the selectivity
site is calculated.  Finally, we consider the electrostatic coupling
between the charge fluctuations in the channel mouth and the energy
fluctuations for the ion at the selectivity site.

It is shown that the volume charge fluctuations in the channel mouth
can be modeled as a generalized short noise and result in the strong
modulation of the potential barrier for an ion at the selectivity
site at the sub-nanosecond time scale. It is further demonstrated
that the enhanced by the channel modulation of the potential barrier
is much larger as compared to the energy of Coulomb interaction
between ion in the channel mouth and the ion at the selectivity
site. We therefore argue that open ionic channels can be viewed as
electrostatic amplifiers of the charge fluctuations. An analytical
approximation to the effect of amplification of the electrostatic
interaction between the ions in the channel is provided.

\section{Method}
\label{s:method}

The system considered is made of three compartments of equal size.
The middle block constitutes the protein through which there is a
cylindrical hole representing an approximation of an open channel.
The channel is bathed by a solution of NaCl on its left and right
ends. The electrostatic force and potential are produced by the
moving ion and fixed charge at the surface, particularly the net
negative charge bared by a fraction of the protein at the channels
selectivity filter. The Poisson equation is solved in 3D space with
cylindrical symmetry. The electrical sources are the fixed negative
charge on the protein and the ions located on the channel axis.
Since the dielectric value of the aqueous pore is not yet known for
narrow channels, as first approximation, we use the same dielectric
value as for the aqueous bath $\epsilon_{2}=80$. The dielectric
value of the protein is taken to be equal to $\epsilon_{1}=2$.

The motion of the ions is modeled within a self-consistent framework
of Brownian dynamics (BD) coupled to the Poisson equation.
\begin{eqnarray}
 \label{eq:poisson_LE}
  m_i \ddot{\vec{x_i}} &=& - m_i \gamma _i \dot{\vec{x_i}}  +
  \left[ {\frac{{q_i q_j }}{{4\pi \varepsilon \varepsilon _0 r_{ij} ^2 }}
   + \frac{{9U_0 R_c^9 }}{{r_{ij} ^{10} }}} \right]
  \frac{{\vec r_{ij} }}{{r_{ij} }} + F_{ch}  + \sqrt {2m_i \gamma _i k_B T} \vec \xi _i (t), \\
  m_j \ddot{\vec{x_j}} &=& - m_j \gamma _j \dot{ \vec{ x_j}}  +
  \left[ {\frac{{q_i q_j }}{{4\pi \varepsilon \varepsilon _0 r_{ji} ^2 }}
   + \frac{{9U_0 R_c^9 }}{{r_{ji} ^{10} }}} \right]
  \frac{{\vec r_{ji} }}{{r_{ji} }} + F_{ch}  + \sqrt {2m_j \gamma _j k_B T} \vec \xi _j (t),
\end{eqnarray}
where $m_i$,~$x_{i}$~and $q_{i}$ are the mass, position and charge
of the $i$th ion.  In the Brownian dynamics simulations, water
molecules are not taken into account explicitly and are
represented in Eq.~(\ref{eq:poisson_LE}), by an average frictional
force with a friction coefficient $m_{i}\gamma_{i}$ and a
stochastic force $\sqrt {2m_i \gamma _i k_B T} \vec \xi _i (t)$
arising from random collisions. The long range Coulomb interaction
is represented by the $1/r$ potential. The addition of the
pairwise repulsive $1/r^9$ soft-core interaction potential insures
that ions of opposite charge, attracted by the inter-ion Coulomb
force, do not collide and annihilate each other. $r_{ij}$ is the
distance between ions $i$ and $j$. $U_{0}$ and $R_{c}$ are
respectively the overall strength of the potential and the contact
distance between ions pairs~\cite{Chung:00a}.

We used the following parameters for the simulations:

Dielectric constants: $\varepsilon_2=80$, $\varepsilon_1=2$;

Masses (in kg): $m_{Na} = 3.8\times10^{-26}$,
$m_{Cl}=5.9\times10^{-26}$;

Diffusion coefficients (in $m^{2}s^{-1}$):
$D_{Na}=1.33\times10^{-9}$, $D_{Cl}=2.03\times10^{-9}$,

(Note that D is related to the friction coefficient via
$D=\frac{k_{B}T}{m\gamma}$);

Ion radii (in \AA): $r_{Na}=0.95$, $r_{Cl}=1.81$;

Temperature: $T = 298$ K.
\begin{table}[!ht]
\begin{center}
\begin{tabular}{|r|c|c|}
\hline
Ions & {\it U$_{0}$(k$_{B}$T)} & {\it R$_{c}$}(\AA) \\
\hline
Na-Na & 0.5 & 3.50 \\
Na-Cl & 8.5 & 2.76 \\
Cl-Cl & 1.4 & 5.22 \\
\hline
\end{tabular}
\caption{Parameters used in the calculation of the short range
ion-ion interaction} \label{tab:short range ion-ion parameters}
\end{center}
\end{table}

The dielectric force acting on the ion as it moves on the channel
axis $F_{ch}$ is calculated numerically by solving Poisson equation
\begin{eqnarray}
\label{eq:poisson_1}
 -\nabla\cdot(\epsilon(\vec{r})\nabla\phi(\vec{r})) =
 \frac{\rho(\vec{r})}{\varepsilon_{0}}, \qquad
 \vec{D} =\epsilon\vec{E}, \qquad
\vec{E}=-\vec{\nabla}\phi
\end{eqnarray}
using finite volume method (FVM)~\cite{Ferziger:96}. In
Eq.~(\ref{eq:poisson_1}), $\epsilon = \varepsilon_{k}
\varepsilon_{0}$ is the space dependent dielectric function with
$k=1$ or $2$. $\varepsilon_{0}=8.85\cdot10^{-12} CV^{-1}m^{-1}$ is
the dielectric constant of empty space. $\rho$ is the source
density, $\phi$ is the potential, $\vec{D}$ and $\vec{E}$ are
respectively the displacement vector and the electric field. An
effective dielectric constant is introduced at the interface between
water and the protein. The later procedure is more appropriate when
the dielectric function present severe jump. In order to preserve
the system symmetrical symmetry, the moving ion is bound to move on
the channel axis. Dirichlet boundary condition is used to fix the
value of the electrostatic potential at the left and right
boundaries of the system as used in experimental measurement.
Neumann boundary condition is used to set the value of the normal
component of the electric field. Standard iterative method is used
to solve algebraic linear system grown from discrete Poisson
equation. The results are stored in tables allowing for the fast
self-consistent BD simulations.

\section{Distribution of arrival times and the charge fluctuations}
\label{s:arrival_times}

In this research we are focused on estimating the effect of the
charge fluctuations on the transition probabilities of the ions
through an open ionic channel. To this end we would like to
calculate how much time ions spend in the channel mouth and how
often do the arrive to the channel mouth. Estimation of arrival time
of ions at the channel mouth from a solution assuming charge
neutrality and no applied field can be obtained by considering pure
diffusion of ions through the hemisphere at the mouth of the channel
and is given by $\tau_{ar}=1/(2\pi z_{i} \aleph D_{i}rc_{0})$; where
$r$ is the channel radius, $c_{0}$ the bulk concentration, $\aleph$
the Avogadro number, $z_{i}$ and $D_{i}$ the charge valence and the
diffusion coefficient of the $i$th ion~\cite{Hille:92,Peskoff:88}.
For an NaCl solution of concentration $c_{0}=500$mM, estimated
arrival time $\tau_{ar}\sim 2.9\times 10^-9$ses for Na$^{+}$ and
$\tau_{ar}\sim 3.8\times 10^-9$ses for Cl$^{-}$.

\begin{figure}[h]
\includegraphics[width=3.2in, height=2.95in]{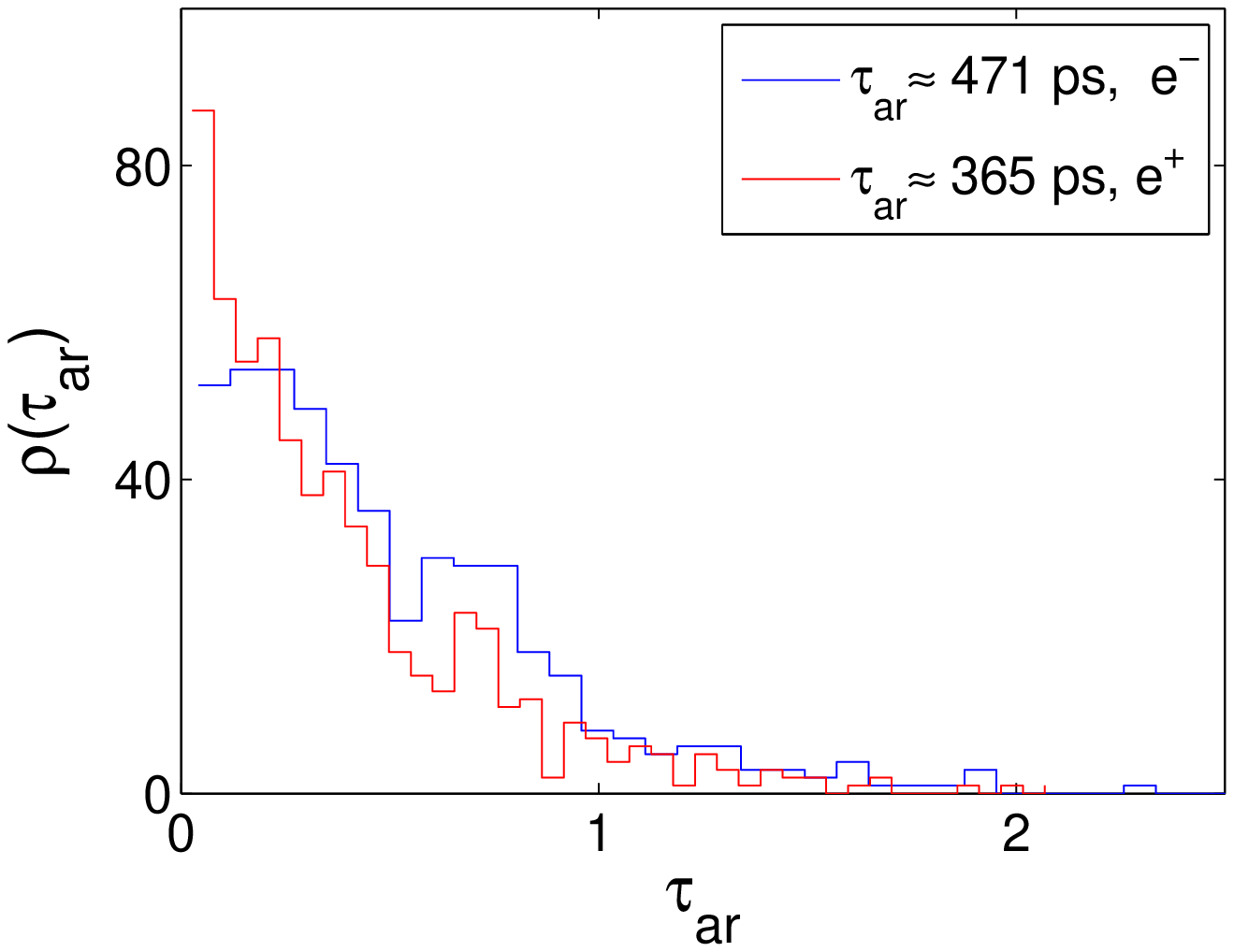}
\includegraphics[width=3.2in, height=2.95in]{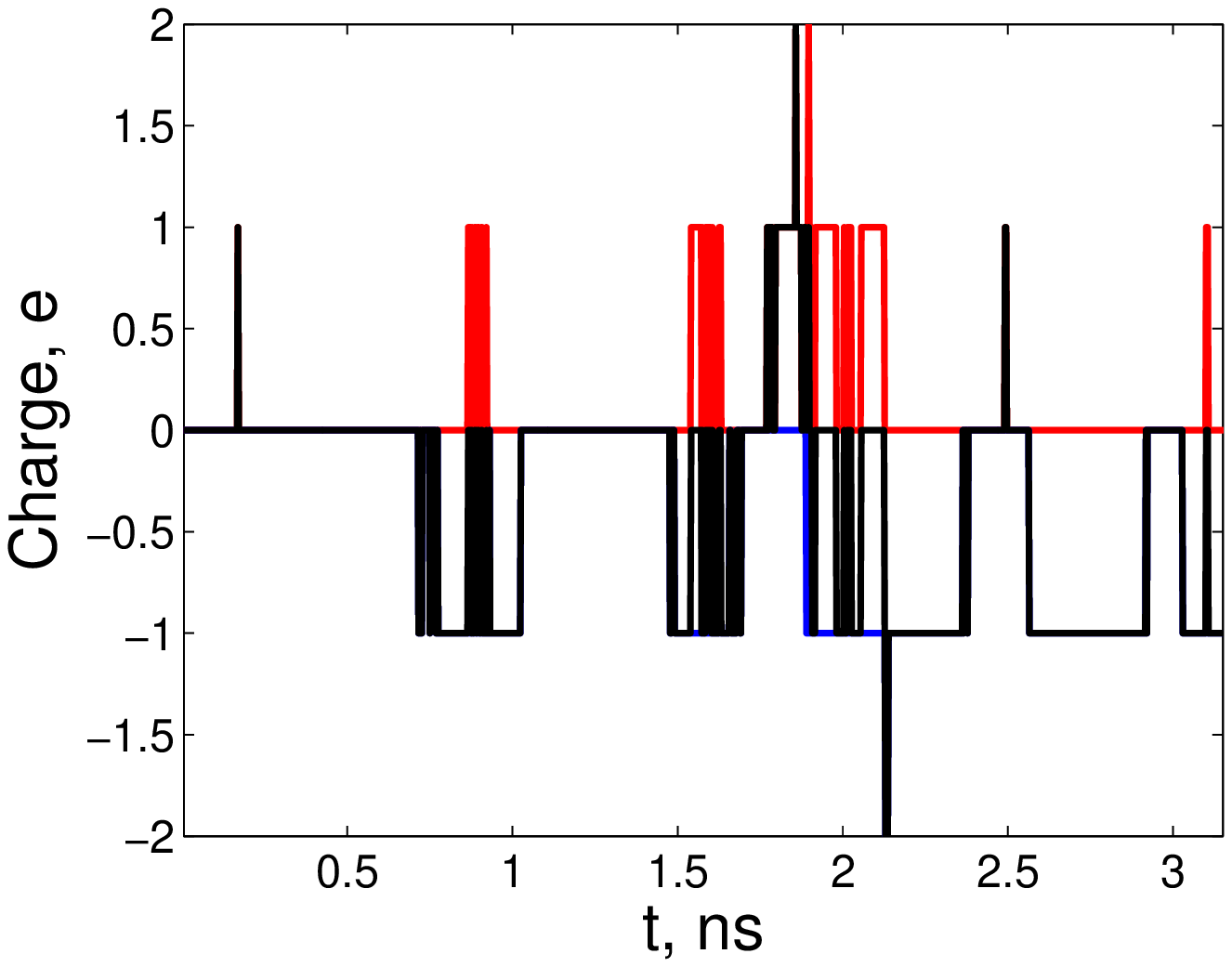}
\caption{\label{fig:arrival}{\bf(a)}  The arrival time
distributions for positive (red line) and negative (blue line)
ions for cylindrical channel of radius r = 4 {\AA}. {\bf(b)}
Fluctuations of the charge at the mouth entrance.}
\end{figure}

To estimate the charge fluctuations from the simulations  we have
recorded continuously (during a few microseconds) the total positive
and negative charge in the channel mouth with volume $v_M=\pi
r^2*r$, where $r=6 \AA$. The arrival time was estimated by recording
the interval of times between the subsequent events of ion arrival
to the channel mouth. The results of simulations are summarized in
the Fig.~\ref{fig:arrival}. It can be seen from the figure that the
arrival time distribution follows exponential distribution with mean
arrival time $\tau_{ar}\sim 3.6\times 10^-9$ses for Na$^{+}$ and
$\tau_{ar}\sim 4.7\times 10^-9$ses for Cl$^{-}$.

This corresponds corresponding to a Poisson process as expected from
the theory. The corresponding charge fluctuations at the channel
mouth is represented on Fig.~\ref{fig:arrival}.{\bf(b)}.

\section{Amplification of the electrostatic interaction between
two ions in the channel}
\label{s:amplification}

We now consider the effect of the charge fluctuations on the
transition probabilities of the ions through an open ionic channel.
First, we analyze the energy profile faced by one ion moving through
the channel.
\begin{figure}[t!]
\includegraphics[width=3.2in, height=2.95in]{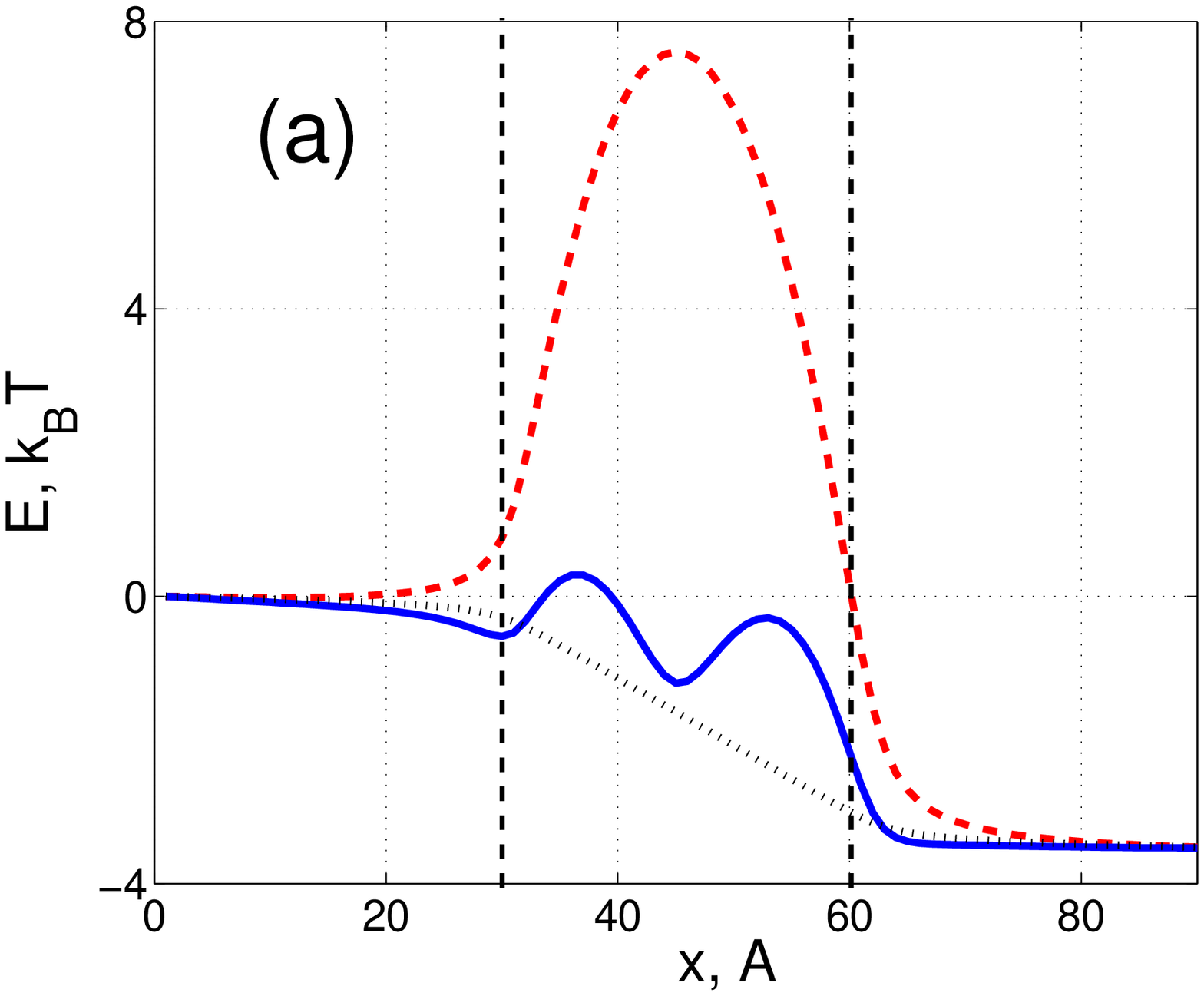}
\includegraphics[width=3.2in, height=2.95in]{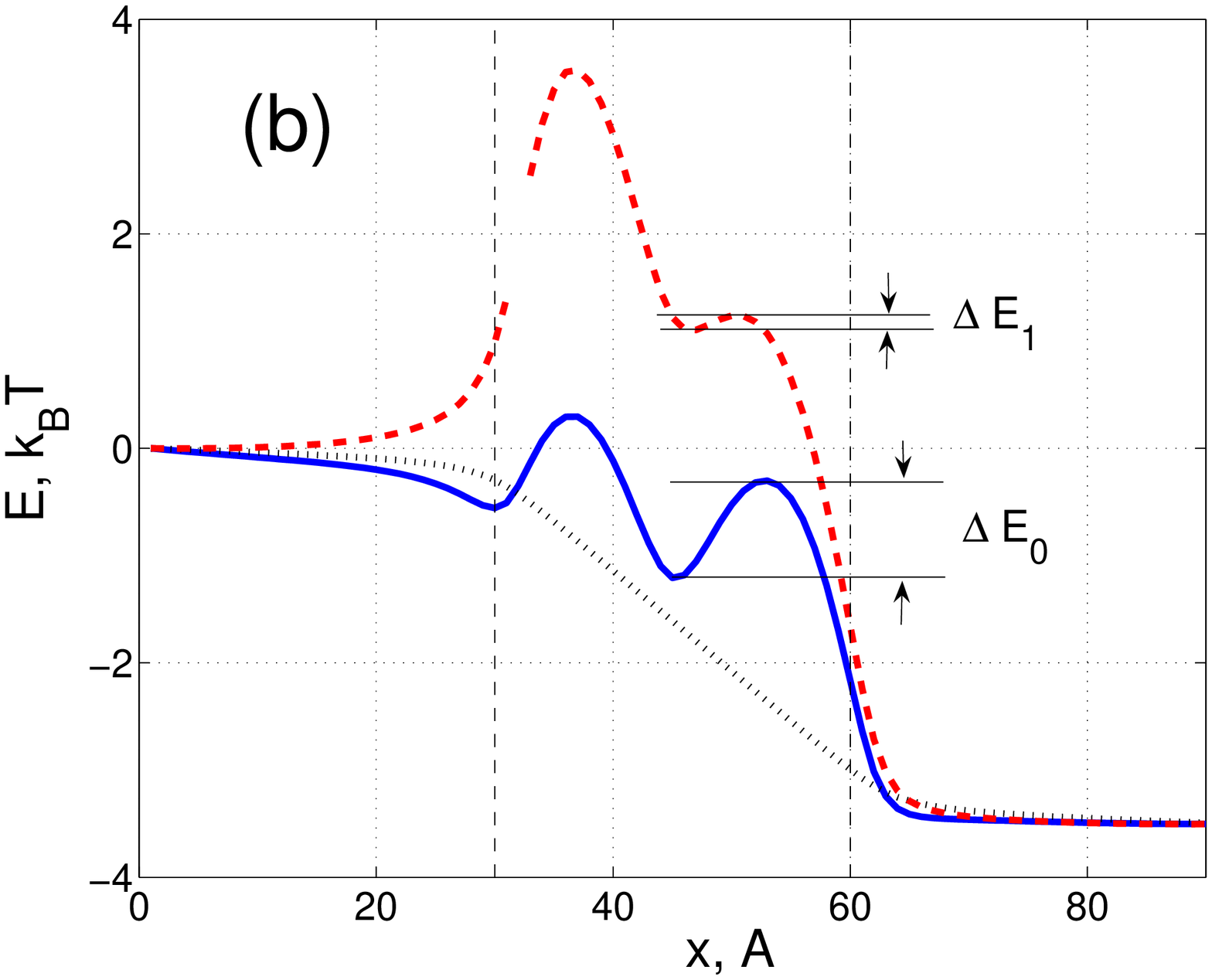}
\caption{\label{fig:energy} {\bf(a)} The potential energy profiles
in a cylindrical channel of radius r = 4 {\AA} when an electric
field of 10$^7$ V/m is applied in the z direction. The solid (blue)
and dashed (red) lines correspond to the channels with and without
fixed charges, respectively. The profile of a passive channel
($\epsilon_{protein}$ = 80) is indicated by the dotted (black) line.
The channel is situated between 30 and 60~\AA. {\bf(b)} Total energy
of the channel as a function of the position of the ion when: the
first ion is located at the channel mouth (dashed line); the channel
is empty (solid line); no channel (dotted line). Vertical
dashed-dotted lines show channel entrance. The height of the
potential barrier for the ions at the selectivity site and no ion at
the channel mouth is denoted $\Delta E_0$. In the presence of the
second ion in the channel mouth this barrier is reduced to $\Delta
E_1$.}
\end{figure}

The total electrostatic energy required to move one ion from the
bulk solution to a bare channel that is $30$~\AA long and $4$~\AA in
radius as a function of the position of the ion is calculated as
shown in the Fig.~\ref{fig:energy}{\bf(a)}. The potential drop
across the channel is $\Delta\Phi$ = 90 mV, the charge at the
selectivity site (at the middle of the channel) is -1e. Note that
the value of the potential barrier for the ion at the selectivity
site to exit the channel is $\Delta E_0$ as shown in the
Fig.~\ref{fig:energy} {\bf(b)}. We now consider the reduction of
this potential barrier induced by the second ion situated at the
channel mouth. The total energy of the channel as a function of the
position of the first ion moving along the channel when the second
ion is located at the channel mouth is shown in the
Fig.~\ref{fig:energy}{\bf(b)} by dashed line. It can be seen that
the presence of a second ion at the channel entrance decreases the
energy barrier to  $\Delta E_1$. We note the reduction of the
potential barrier from  $\Delta E_0$ to $\Delta E_1$ is much
stronger in the channel with low dielectric constant.  I.e. the
effect of charge fluctuation is strongly enhanced by the protein low
dielectric constant compared to water. in the absence of the protein
walls the interaction between two ions will be purely due to the
Coulomb forces and the corresponding reduction of the potential
barrier $\Delta E_C = \frac{e^2}{4\pi\epsilon_2\epsilon_0 r_{12}}$
will be much smaller then the effect induced by the channel $\Delta
E_0-\Delta E_1$.

\section{Analytical approximation of the effect of the amplification}
\label{s:analytics}

A simple one-dimensional approximation of Poisson equation for a
long (length $d$) and narrow (radius $r$) channel can be
derived~\cite{Barcilon:92a,Barcilon:92b} and written on its
dimensionless form as:
\begin{eqnarray}
\label{eq:pois an_9} \ddot{\Phi}-\beta\Phi &=& -\beta\Delta(1-x)
-\alpha\rm P(x) -\alpha\delta(x-x_{0})-\alpha\delta(x-x_{1})\\
\Phi(0) &=& 0, \;\;\;\; \Phi(1) = \Delta \nonumber
\end{eqnarray}
where $\beta=\frac{d^{2}\tilde{\varepsilon}}{\varepsilon _0
\varepsilon _{H_2 O}}$ is a function of the effective dielectric
parameter $ \tilde \varepsilon  = \frac{{\varepsilon _p
}}{{\varepsilon _{H_2 O} }}\frac{{2\varepsilon _0 }}{{r^2 \ln
\left( {{d \mathord{\left/
 {\vphantom {d r}} \right.
 \kern-\nulldelimiterspace} r}} \right)}}$. The factor
$\alpha=\kappa^{2}d^{2}$ where
$\kappa^{-1}=\sqrt{\frac{\varepsilon _0 \varepsilon _{H_2 O}
v_{0}U_{T}}{e}}$~is the Debye length with $v_{0}=\pi r^{2}d$; the
volume of the cylindrical channel. $U_{T}=k_{B}T$ is the thermal
energy; where $k_{B}$ is the Boltzmann's constant and $T$ the
absolute room temperature. $P(x)$ is the permanent charge
distribution on the protein atoms; independent of the electric
field. The $\delta$ function represents two moving ions at
different locations $x_{0}$ and $x_{1}$ on the channel axis.
Because the Poisson equation is linear the contribution from the
four terms in the right hand side can be considered independently.
Here we concentrate on the interaction between two ions in the
channel in the absence of any external field and therefore leave
only the two last terms in the eq. (\ref{eq:pois an_9}) (see,
however, for the full analytical solution of the Poisson equation
elsewhere).

\begin{eqnarray}
\label{eq:pois an_10} & &\Phi(x)= -\ \alpha\int_{0}^{1}G(x,s)P(s)ds \nonumber\\
& &-\frac{\alpha}{\sqrt{\beta}\sinh\sqrt{\beta}}\left\{
\begin{array}{c}
   \sinh\sqrt{\beta}x\sinh\sqrt{\beta}(x_{0}-1)+\sinh\sqrt{\beta}x\sinh\sqrt{\beta}(x_{1}-1),\;\;0\leq x\leq x_{1}\\
   \sinh\sqrt{\beta}x_{0}\sinh\sqrt{\beta}(x-1)+\sinh\sqrt{\beta}x\sinh\sqrt{\beta}(x_{1}-1),\;\;x_{1}\leq x\leq x_{0}\\
   \sinh\sqrt{\beta}x_{0}\sinh\sqrt{\beta}(x-1)+\sinh\sqrt{\beta}x_{1}\sinh\sqrt{\beta}(x-1),\;\;x_{0}\leq x\leq 1\\
\end{array}
\right.\nonumber\\
\end{eqnarray}
Where the Green function is given by:
\begin{eqnarray}
\label{eq:pois an_5}G(x,s)&=&
\frac{1}{\sqrt{\beta}\sinh\sqrt{\beta}}\left\{
\begin{array}{c}
    \sinh\sqrt{\beta}x\sinh\sqrt{\beta}(s-1),\;\;\;\;0\leq x\leq s \\
    \sinh\sqrt{\beta}s\sinh\sqrt{\beta}(x-1),\;\;\;\;s\leq x\leq 1 \\
\end{array}
\right.
\end{eqnarray}
The distribution of the permanent charge is modelled using a
narrow Gaussian distribution, consistent with the fact that the
charge is mainly concentrated at the central part of the channel
\[
P(x)=\frac{1}{\sqrt{2\pi\sigma^{2}}}exp(-\frac{(x-\mu)^{2}}{2\sigma^{2}})
\]

\section{Conclusion}
\label{s:conclusions}

It was shown that the energy fluctuations due to the correlation of
the motion between ions in strong electrolytes can be substantially
enhanced in the open ionic channels with low dielectric constant.
Therefore, ion channels can be taught of as electrostatic amplifiers
of the charge fluctuations. This in turn may lead to an
exponentially strong modulation of the potential barrier for the ion
at the selectivity site. The resulting enhancement of the transition
probabilities of ions through open ionic channels can be leading
order contribution to the transition probabilities calculated using
standard theories that neglect the effect of charge fluctuations.
The effect of electrostatic amplification of the charge fluctuations
may play substantial role in the nanoscale protein dynamics,
including, in particular, protein folding.


\section{References}
\begin{thebibliography}{10}

\bibitem{Barcilon:92a}
{\sc V.~Barcilon}, {\em Ion flow through narrow membrane channels
.1}, SIAM J.
  Appl. Math., 52 (1992), pp.~1391--1404.

\bibitem{Barcilon:92b}
{\sc V.~Barcilon, D.~P. Chen, and R.~Eisenberg}, {\em Ion flow
through narrow
  membrane channels .2}, SIAM J. Appl. Math., 52 (1992), pp.~1405--1425.

\bibitem{Chung:98}
{\sc S.~H. Chung, T.~W. Allen, M.~Hoyles, and S.~Kuyucak}, {\em
Permeation of
  ions across the potassium channel: Brownian dynamics studies}, Biophys. J.,
  77 (1998), pp.~2517--2533.

\bibitem{Chung:99}
{\sc S.~H. Chung, M.~Hoyles, T.~W. Allen, and S.~Kuyucak}, {\em
Study of ionic
  currents across a model membrane channel using brownian dynamics}, Biophys.
  J., 75 (1999), pp.~793--809.

\bibitem{Ferziger:96}
{\sc J.~Ferziger and M.~Peric}, {\em Computational Method for Fluid
Dynamics.},
  Springer, Berlin, 1996.

\bibitem{Hille:92}
{\sc B.~Hille}, {\em Ionic Channel Of Excitable Membranes}, Sinauer
Associates,
  Sunderland, MA, 1992.

\bibitem{Landau:80a}
{\sc L.~D. Landau and E.~M. Lifshitz}, {\em Statistical Physics},
vol.~5 of
  Course of Theoretical Physics, Pergamon, Oxford, 3~ed., 1980.
\newblock Part 1.

\bibitem{Chung:00a}
{\sc G.~Moy, B.~Corry, S.~Kuyucak, and S.-H. Chung}, {\em Tests of
continuum
  theories as models of ion channels. i. poisson-boltzmann theory versus
  brownian dynamics}, Biophys. J., 78 (2000), pp.~2349--2363.

\bibitem{Debye:23}
{\sc D.~P and H.~E}, {\em The theory of electrolytes. i. lowering of
freezing
  point and related phenomena}, Physik. Z., 24 (1923), pp.~185--206.

\end{thebibliography}

\end{document}